\documentclass [11pt]{article}
\usepackage[dvips]{color}
\usepackage{epsfig}
\usepackage[normalem]{ulem}
\def\beq{\begin{equation}}
\def\eeq{\end{equation}}
\def\beqa{\begin{eqnarray}}
\def\eeqa{\end{eqnarray}}

\begin{document}
\baselineskip0.6cm plus 1pt minus 1pt
\tolerance=1500

\begin{center}
{\LARGE\bf ``Walking" along a free rotating bicycle wheel \\
(Round and round)   }
\vskip0.4cm
{ J. G\"u\'emez$^{a,}$\footnote{guemezj@unican.es},
M. Fiolhais$^{b,}$\footnote{tmanuel@teor.fis.uc.pt}
}
\vskip0.1cm
{\it $^a$ Departamento de F\'{\i}sica Aplicada}\\ {\it Universidad de
Cantabria} \\ {\it E-39005 Santander, Spain} \\
\vskip0.1cm
{\it $^b$ Departamento de F\'\i sica and Centro de
F\'\i sica Computacional}
\\ {\it Universidade de Coimbra}
\\ {\it P-3004-516 Coimbra, Portugal}
\end{center}

\begin{figure}[ht]
\begin{center}
\includegraphics[width=5cm]{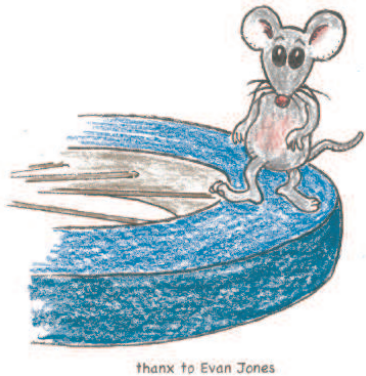}
\end{center}
\vspace*{-0.8cm}
\caption[]{\label{fig:1} \small Marta mouse ``walking" along a free rotating bicycle wheel as shown in \cite{hewitt2014a}.}  
\end{figure}

We describe the kinematics, dynamics and also some energetic issues related to the Marta mouse motion when she walks on top of a horizontal bicycle wheel, which is free to rotate like a merry-to-go round, as presented recently by Paul Hewitt in the Figuring Physics section of this magazine.  The situation is represented in figure
\ref{fig:1}, which was taken from ref. \cite{hewitt2014a}.

Physics of walking contains some interesting features that are undervalued (or, at least, not always stressed) by physics teachers and even by authors of textbooks. One of those features is related to the forces in a stride. Indeed, it is common to mention that,
when one walks, the horizontal force on the foot points in the forward direction. This is definitely true in a certain time interval. But, in another time interval, during the very same stride, the horizontal force points backwards. We do not know of any textbook where such a time dependent horizontal force, pointing alternately  forward and backwards, is discussed in a walking context. On the other hand, energetic issues in walking are usually not discussed either.

We address these points in the present article, which was inspired by the quiz presented in \cite{hewitt2014a}.  In brief, the question in \cite{hewitt2014a} is the following: Marta mouse, initially at rest, starts walking along the outer surface of a bicycle wheel, placed horizontally, initially also at rest. When Marta stops, what happens, kinematically speaking, to the wheel? The answer is unquestionable: the wheel also stops \cite{hewitt2014b}. If, in the initial state, everything is at rest, as a consequence  of the conservation of the angular momentum that is always zero, the wheel should come to rest when Marta stops. The conservation of the angular momentum is a consequence of the vanishing external torque (the torque of the internal forces is zero).

The quiz is only about the initial and the final states, but we think it is interesting to describe, even in some qualitative way, the whole process: from the initial rest state until the final rest state of both Marta and the wheel. But, before analyzing the motion on the wheel, it is instructive to considerer the simpler case of a normal walking. To start walking (this applies to a person, to an animal, to a toy, etc.), the foot must exert a backward force on the ground. According to Newton's third law, the ground exerts an opposite force, i.e. a forward force, on that foot. The body is already moving when the other foot now touches the ground. This other foot is moving forward with respect to the ground, so the static friction force exerted by the ground on the foot is initially opposite to that direction: the force points backwards. However, the foot is articulated and it starts to exert a backward force on the ground and then it experiences a forward force as a result, again, of Newton's third law. And the process repeats cyclically. In \cite{guemez2013c} we presented a simple model of constant forces to describe the horizontal force exerted on the feet of a walking person.

In spite of the fact that Marta is walking along a circular trajectory, the description of her motion is along the same lines.
When Marta begins to walk along the surface, her foot pushes the tire backwards with respect to her motion, causing the wheel to rotate. The third law of mechanics implies that the wheel exerts an opposite (i.e. a forward) force on Marta's foot, and therefore she acquires a forward (tangential) acceleration.

Since the trajectory of Marta is a circle with respect to a fixed inertial reference frame whose origin is the center of the wheel, the tire must also exert a centripetal force on Marta (a slight inclination inwards would help Marta). But, let's concentrate only on the tangential  force on Marta's feet. When the other foot, for the second stride, touches the wheel, the tangential force exerted by the tire points initially backwards, causing a reduction of Marta's center-of-mass linear velocity. However, shortly afterwards, and on the same foot, the force is again in the forward direction. Therefore the next stride starts with a deceleration period followed by an acceleration period and typically, after a few steps, a ``cruising walking speed", which actually is never a constant speed, is achieved.

The stopping process is the inverse of the starting one, in the sense that, in the last strides, the backward force should now produce an impulse that, in absolute value, is larger than the impulse of the previous forward force.

\begin{figure}[thb]
\begin{center}
\includegraphics[width=10cm]{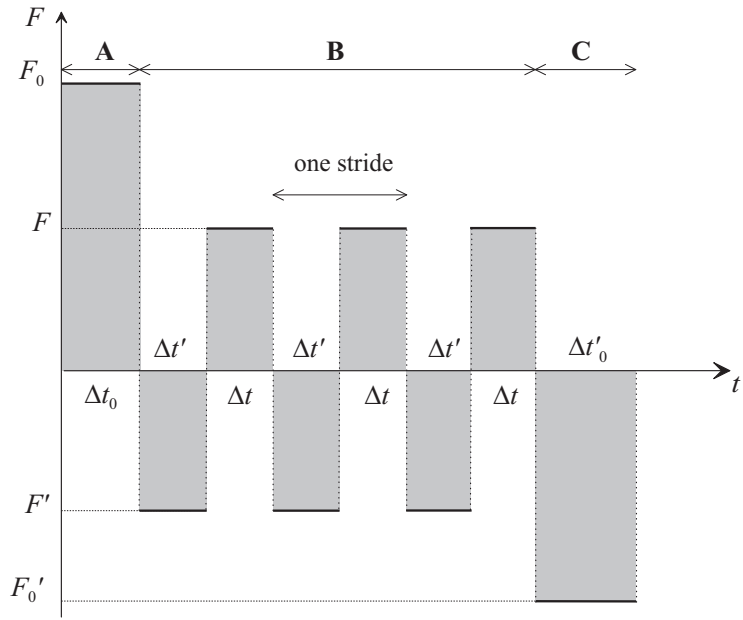}
\end{center}
\vspace*{-0.8cm}
\caption[]{\label{fig:2} \small Force acting on the feet of Marta when she walks along the bicycle wheel. {\bf A} corresponds to the starting phase, {\bf B} to the ``cruising speed" phase and {\bf C} to the stopping phase. } 
\end{figure}

The same simple model of constant forces used in \cite{guemez2013c} to describe the horizontal force on the feet of a walking person  is now applied to the time dependent tangential force on Marta. This means that the actual horizontal forces on Marta are approximated by those represented in figure \ref{fig:2}. The model is oversimplified because: i) the actual forces are not constant; ii) the initial and final phases comprise more than one step each, though we assume just one step for each of those phases. But these are mere details that do not question the global description of the process in our schematic model. We do not require that the positive and the negative forces, acting during the ``cruising speed" regime, are of the same magnitude, although in figure \ref{fig:2} we chose to have $\Delta t=\Delta t'$ and therefore $F=-F'$. The important point is that their impulses cancel each other, so that the magnitude of Marta's center-of-mass velocity does not change after a stride (except for the first and for the last one). On the other hand, in figure \ref{fig:2} we choose to have $\Delta t'_0>\Delta t_0$, so that $F_0>-F'_0$.
The algebraic sum of the shadowed areas (the impulses of the force) in figure \ref{fig:2} should add up to zero. Since the torque is given by force times the radius $R$ of the wheel, what we just stated for the impulse of the tangential force also applies to the angular impulse given by $FR\Delta t$, where $F$ and $\Delta t$ here denote, in a  generic way, the force and the time interval.

\begin{figure}[htb]
\begin{center}
\includegraphics[width=10cm]{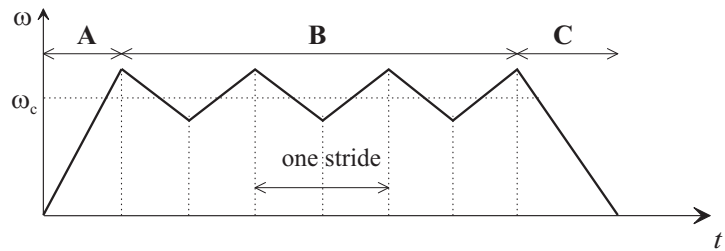}
\end{center}
\vspace*{-0.8cm}
\caption[]{\label{fig:3} \small Angular velocity of Marta as a function of time. By $\omega_{\rm c}$ we denote the ``cruising angular velocity".} 
\end{figure}

This simple model leads to a time dependent linearly varying  velocity, $v(t)$. The angular velocity of Marta with respect to the rotation axis is $\omega=v/R$, which is also a time dependent function --- see figure \ref{fig:3}. As the figure shows, there is no time interval during which the angular velocity of Marta is constant, but we may still define a ``cruising angular velocity", $\omega_{\rm c}$. The qualitative picture provided by figure \ref{fig:3} is somehow exaggerated in the sense that the actual deviations from  $\omega_{\rm c}$, in the steady state regime ({\bf B}), are supposedly smaller.

If one assumes that Marta's mass, $m$, is rather small, the center of mass of the system practically still lies at the center of the wheel. Her moment of inertia, $I=mR^2$, is
 then much smaller than the moment of inertia of the wheel, $I_0 \gg I$. The angular velocity of the wheel, $\Omega$, is also small: from the conservation of the angular momentum, that is always zero, one immediately concludes that $\Omega(t)=-\omega(t) {I \over I_0}$ (the minus sign indicates that the rotation is in the opposite direction).

\begin{figure}[ht]
\begin{center}
\includegraphics[width=6.5cm]{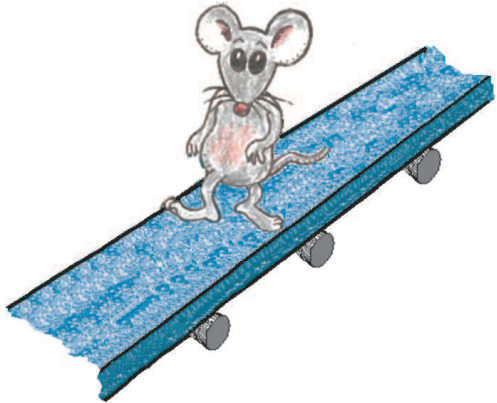}
\end{center}
\vspace*{-0.5cm}
\caption[]{\label{fig:4}  \small Marta mouse ``walking" along a free sliding horizontal board.}  
\end{figure}

The present situation is very similar to the well-known example of an isolated system of two bodies of masses $m$ and $M_0\gg m$ initially at rest, and it is worth stressing this analogy in the classroom. If the first body starts to move with velocity $v(t)$ measured in an inertial reference frame, the other one also moves with velocity $V(t)=-v(t) {m \over M_0}$ (the minus sign indicates that the second body moves in the opposite direction with respect to the first body). If one body stops, the other also has to stop.  The situation is represented in figure \ref{fig:4} again with Marta, now walking along a straight line trajectory on top of a free sliding board. The above discussion presented in the context of the circular motion does apply, {\em mutatis mutandis}, to the situation represented in figure \ref{fig:4}. In particular, the counterpart of the plot shown in figure \ref{fig:3} is a similar plot now for the center of mass velocity \cite{guemez2013c}.

Let's go back to  Marta's rotational motion to discuss other issues. The
braking force $F'_{0}$ in phase {\bf C} (note again that the
 braking process may take place in various steps and not in just one step as assumed in figure \ref{fig:2}) is not necessarily equal and opposite to the force $F_0$ in phase {\bf A} (also note again that the acceleration phase may also take place in various steps and not in just one). These forces are unrelated except for the fact that their impulses (or the corresponding angular impulses) are symmetric as it is the case in figure \ref{fig:2}, since we choose to have all intermediate impulses (phase {\bf B}) adding up to zero. In figure \ref{fig:2}, $F_0\not= -F'_0$, but $F_0\Delta t_0+ F'_0\Delta t'_0=0$, where $\Delta t_0$ and $\Delta t'_0$ are the initial and final time intervals as represented in the figure.

 As mentioned above, Newton's third law also plays a role here  (but not in the way expressed in \cite{hewitt2014b}). Marta exerts a tangential force on the wheel (or, better to say, a force with a tangential component) that is, at any time, equal and opposite to the force that the wheel exerts on her.
The force   on the wheel is responsible  for  its rotation with angular velocity $\Omega(t)$.

The radial component of the contact force is responsible for the centripetal acceleration on Marta, of magnitude $R\omega^2$. Similarly, a force pointing in the opposite direction is exerted by Marta on the wheel (Newton's third law). The study of these forces is also straightforward, though in this paper we are concentrated mainly on the tangential force. On the other hand, if the mass of Marta is reasonably large, the problem gets a bit more complicated, but still manageable, because the system center-of-mass is no longer at the center of the wheel. However, the study of that problem is out of the scope of the present note.

The model we have used for the tangential force is a simple one but good enough. Had we used a more realistic one, for instance by assuming a time dependent force of sinusoidal shape, as measured in humans \cite{haugland2013}, the consequence on the angular velocity of Marta would be a smoothing out of the edges in figure \ref{fig:3}, corresponding to a continuous angular acceleration function. Nevertheless, and in spite of its simplicity, our model accounts for the most fundamental points.

To conclude, a brief reference to some energetic issues. In a recent publication \cite{guemez2014b} we studied the energy transfers  in various human movements such as whirling, jumping and walking. The arguments presented in \cite{guemez2014b} also apply here. The motion is possible because Marta possesses an internal source of energy. If Marta is a living being, the origin of the energy lies in the biochemical reactions ($\xi$) that take part in her muscles. If Marta is a toy, the internal energy can be provided by a spiral spring (as it is the case of the kangaroo toy of ref \cite{guemez2014c}) or just by batteries.

The first law of thermodynamics is the pertinent physical law to discuss the energetic issues. In general, that law can be  expressed by~\cite{guemez2013a} $\Delta K_{\rm cm} + \Delta U =
W_{\rm ext} + Q$, where, on the left-hand side, we
identify the variation of the center-of-mass kinetic energy and the variation of the system's internal
energy due to all possible reasons (in the present case it includes $\Delta U_\xi$ for the biochemical reactions and also the variations of the rotational kinetic energies ${1\over 2} I \omega^2$ and ${1\over 2} I_0 \Omega^2$ \cite{mallin92}) and, on the right-hand side, we identify the work performed by the
external forces, and the other possible energy transfers to/from the system that are not work, so they are
heat. Defining Marta and the wheel as the thermodynamical system and considering the global process,  from the initial rest state to the final rest state, the variation of the center-of-mass kinetic energy is obviously zero. The variation of the rotational kinetic energies is also zero, so that the left hand side reduces to $\Delta U_\xi$  (note also that the internal forces applied by Marta on the wheel and vice-versa do not perform any work).
The external forces are the weights and the force on the wheel axis that do not do any work either.   Therefore the expression of the first law of thermodynamics, considering the global process, reduces to
$\Delta U_\xi=Q$, which is a negative quantity because there must be a reduction of the internal energy.

The conclusion is clear: the internal energy provided by Marta is transferred as heat to the surrounding where it is dissipated, and the global process leads to an entropy increase of the universe \cite{guemez2013c,guemez2014b,guemez2014c}. Certainly, part of this energy can also be reabsorbed by Marta but this energy degradates, in the sense that it will not be useful anymore (for, instance, the temperature of Marta may slightly increase). In fact,   one should note that the energetic asymmetry studied in \cite{guemez2014b}, in relation with the two phases of the stride, also shows up here: in the acceleration phase of the step, there is a conversion of internal energy into kinetic energy of the system; in the
braking phase, part of the kinetic energy is dissipated as heat that flows to the surrounding. This is similar to the energy transfers in the normal walking process \cite{youtube}.

\end{document}